\documentclass[english,aip,jcp,preprint]{revtex4-1}
\usepackage[T1]{fontenc}
\usepackage[latin9]{inputenc}
\usepackage{babel}
\usepackage{textcomp}
\usepackage{bm}
\usepackage{amsmath}
\usepackage{amssymb}
\usepackage{graphicx}
\usepackage{esint}
\usepackage[unicode=true,pdfusetitle,
 bookmarks=true,bookmarksnumbered=false,bookmarksopen=false,
 breaklinks=false,pdfborder={0 0 1},backref=false,colorlinks=false]
 {hyperref}

\makeatletter

\providecommand{\tabularnewline}{\\}

\@ifundefined{textcolor}{}
{%
 \definecolor{BLACK}{gray}{0}
 \definecolor{WHITE}{gray}{1}
 \definecolor{RED}{rgb}{1,0,0}
 \definecolor{GREEN}{rgb}{0,1,0}
 \definecolor{BLUE}{rgb}{0,0,1}
 \definecolor{CYAN}{cmyk}{1,0,0,0}
 \definecolor{MAGENTA}{cmyk}{0,1,0,0}
 \definecolor{YELLOW}{cmyk}{0,0,1,0}
}

\makeatother

\begin{document}
\global\long\def\nofr{n(\boldsymbol{{r}})}
\global\long\def\Pofr{\boldsymbol{P}(\boldsymbol{r})}
\global\long\def\rhoofrOmega{\rho(\boldsymbol{r},\boldsymbol{\Omega})}
\global\long\def\rhob{\rho_{\mathrm{B}}}
\global\long\def\nb{n_{\mathrm{B}}}

\global\long\def\dr{\mathrm{d\boldsymbol{r}}}
\global\long\def\dOmega{\mathrm{d\boldsymbol{\Omega}}}

\global\long\def\calF{{\cal F}}
\global\long\def\Fexc{{\cal F_{\mathrm{exc}}^{\ }}}
\global\long\def\Fid{F_{\mathrm{id}}^{\ }}
\global\long\def\Fext{{\cal F_{\mathrm{ext}}^{\ }}}
\global\long\def\Fcor{{\cal F_{\mathrm{b}}^{\ }}}
\global\long\def\Fthree{{\cal F_{\mathrm{3B}}^{\ }}}
\global\long\def\FexcHNC{{\cal F_{\mathrm{exc}}^{\mathrm{HNC}}}}

\global\long\def\kbT{\mathrm{k_{B}}T}
\global\long\def\rmax{r_{\mathrm{max}}}

\title{Molecular Density Functional Theory for water with liquid-gas coexistence
and correct pressure}

\author{Guillaume Jeanmairet}

\email{g.jeanmairet@fkf.mpg.de}

\affiliation{\'Ecole Normale Sup\'erieure - PSL Research University, D\'epartement de Chimie,  Sorbonne Universit\'es - UPMC Univ Paris 06, CNRS UMR 8640 PASTEUR, 24 rue Lhomond, 75005 Paris, France.}

\author{Maximilien Levesque}

\email{maximilien.levesque@ens.fr}

\affiliation{\'Ecole Normale Sup\'erieure - PSL Research University, D\'epartement de Chimie,  Sorbonne Universit\'es - UPMC Univ Paris 06, CNRS UMR 8640 PASTEUR, 24 rue Lhomond, 75005 Paris, France.}

\author{Volodymyr Sergiievskyi}

\affiliation{SIS2M, LIONS, CEA, Saclay, France}

\author{Daniel Borgis}

\affiliation{\'Ecole Normale Sup\'erieure - PSL Research University, D\'epartement de Chimie,  Sorbonne Universit\'es - UPMC Univ Paris 06, CNRS UMR 8640 PASTEUR, 24 rue Lhomond, 75005 Paris, France.}
\affiliation{Maison de la Simulation, USR 3441, CEA - CNRS - INRIA - Univ. Paris-Sud - Univ. de Versailles, 91191, Gif-sur-Yvette Cedex, France}

\begin{abstract}
The solvation of hydrophobic solutes in water is special because liquid
and gas are almost at coexistence. In the common hypernetted chain
approximation to integral equations, or equivalently in the homogenous
reference fluid of molecular density functional theory, coexistence
is not taken into account. Hydration structures and energies of nanometer-scale
hydrophobic solutes are thus incorrect. In this article, we propose
a bridge functional that corrects this thermodynamic inconsistency
by introducing a metastable gas phase for the homogeneous solvent.
We show how this can be done by a third order expansion of the functional
around the bulk liquid density that imposes the right pressure and
the correct second order derivatives. Although this theory is not
limited to water, we apply it to study hydrophobic solvation in water
at room temperature and pressure and compare the results to all-atom
simulations. The solvation free energy of small molecular solutes like $n$-alkanes and hard sphere solutes whose radii range from angstroms to nanometers is now in quantitative agreement with reference all atom simulations.
The macroscopic liquid-gas surface tension predicted by the theory
is comparable to experiments. This theory gives an alternative to
the empirical hard sphere bridge correction used so far by several
authors.
\end{abstract}
\maketitle

\section{Introduction}

Implicit solvation techniques based on liquid-state theory such as
integral equation theory in the interaction-site\cite{chandler_optimized_1972,hirata_application_1982}
or molecular picture\cite{blum_invariant_1972,blum_invariant_1972-1}
or classical density functional theory\cite{evans_nature_1979,evans_density_2009,henderson_fundamentals_1992} have
proven to be successful for the computation of solvation properties.
Those methods have shown to give thermodynamic and structural results
that get closer and closer to all-atom simulations at a much lower
numerical cost. A current challenge lies in the development and implementations
of three-dimensional implicit solvation theories to describe molecular
liquids and solutions. Recent developments in this direction have
focused on Gaussian field\cite{varilly_improved_2011} theoretical
approaches, or the 3D reference interaction site model (3D-RISM),\cite{beglov_integral_1997,hirata_molecular_2003}
an appealing integral equation theory that has proven recently to
be applicable to, e.g., structure prediction in complex biomolecular
systems. Integral equations are, however, restricted by the choice
of a closure relation, typically, Hypernetted Chain (HNC), Percus-Yevick
or Kovalenko-Hirata. Despite their great potential, they remain difficult
to control and improve, especially for arbitrary three-dimensional
molecules, and they can prove difficult to converge.

We have proposed recently a three dimensional formulation of molecular
density functional theory (MDFT) in the homogeneous reference fluid
approximation (HRF) to study solvation\cite{ramirez_density_2002,jeanmairet_molecular_2013-1}.
It has proven successful in studying solvation properties of solutes
of arbitrary three-dimensional complexity embedded in various molecular
solvents. However, when one comes to water, the HRF approximation
fails even qualitatively to predict the solvation of large hydrophobic
solutes\cite{jeanmairet_molecular_2013}. Such limitation can be explained
by two essential features of water at ambient conditions that are
not properly described by HRF functional. First, it is known that
the solvation free energy of mesoscale apolar solutes can be modeled
as the sum of a surface and volume term\cite{reiss_statistical_1959}.
For water, as well as for any solvent at room condition, the pressure
is very low, inducing negligible $PV$ term until large
radii\cite{huang_scaling_2001}. Another key feature is that at ambient
condition, water is close to its liquid-gas coexistence. As a consequence
the solvation of big hydrophobic solutes may induce dewetting~\cite{huang_hydrophobic_2002}.

In section \ref{sec:Theory}, we propose an extension of MDFT to introduce
the liquid-gas coexistence of the solvent and to recover the correct
pressure. Then, in section \ref{sec:Results-and-Discussion}, we apply
our theory to a model of water and compare the results of solvation
of apolar solutes with reference all-atom Monte Carlo simulations
(MC).

\section{\label{sec:Theory}Theory}

While the theory discussed here is generic to any classical density
functional theory, it is described below in the framework of the MDFT
for water introduced recently\cite{jeanmairet_molecular_2013-1,jeanmairet_molecular_2013}.
We start from the single point charge extended (SPC/E) model of water\cite{berendsen_missing_1987}, that is, a model
comprising one Lennard-Jones site and 3 partial charges. With MDFT,
one computes the solvation free energy and the solvation structure
of a solute of arbitrary shape that acts on the water density field
through an external potential. This last quantity is the sum of an
electrostatic vector field $\boldsymbol{E}(\boldsymbol{r})$ and a
Lennard-Jones scalar field $\Phi_{\mathrm{LJ}}(\boldsymbol{r})$\cite{jeanmairet_molecular_2013-1}.
In the general case, the functional of the density $\rhoofrOmega$
depends upon the position $\boldsymbol{r}$, and the molecular orientation
of the (rigid) solvent molecule. There is no restriction on the molecular
or chemical nature of the solvent molecule model, but to be rigid.
In that particular model of water, $\rhoofrOmega$ can be split into
two distinct fields: the molecular density field $\nofr$ coupled
to $\Phi_{\mathrm{LJ}}(\boldsymbol{r})$ and the polarization vector
field $\Pofr$ coupled to $\boldsymbol{E}\left(\boldsymbol{r}\right)$.
These fields are themselves functionals of $\rhoofrOmega$: 
\begin{equation}
\nofr=\int\rhoofrOmega\dOmega,\label{eq:n(r)=00003DintrhodOmega}
\end{equation}
\begin{equation}
\Pofr=\iint\rho(\boldsymbol{r}^{\prime},\boldsymbol{\Omega})\boldsymbol{\mu}(\bm{r}-\bm{r}^{\prime})\dr^{\prime}\dOmega.\label{eq:P(r)=00003DintrhomudOmega}
\end{equation}
where $\dOmega$ denotes the integration over all molecular orientations.
$\boldsymbol{\mu}(\boldsymbol{r},\boldsymbol{\Omega})\equiv\sum_{\text{m}}q_{\text{m}}\bm{s}_{\text{m}}(\bm{\Omega})\intop_{0}^{1}\delta(\bm{r}-u\bm{s}_{\text{m}}(\bm{\Omega}))\text{d}u$
is the molecular polarization of a single water molecule at the origin
of a cartesian frame, $q_{\mathrm{m}}$ and $\bm{s}_{\text{m}}$ are
the charge and position of the m$^{\mathrm{th}}$ solvent site. One
should refers to Jeanmairet \textit{et al}\cite{jeanmairet_molecular_2013-1}
for a complete description of MDFT for water. Without loss of generality,
we will stick in what follows to solutes without partial charges,
so that the polarization vector field is zero $\left(\boldsymbol{P}=\boldsymbol{0}\right)$.
As a consequence, the free energy is, at dominant order, a functional
of $\nofr$ only.

We now write the Helmholtz free energy functional, $\calF[n]$, that
is the difference of the grand potential of the system containing
the solute, $\Theta$, and without the solute, $\Theta_{\mathrm{B}}$.
In this last case, the solvent is homogeneous at density $n_{\textrm{B}}$
(typically 1~g/cm$^{3}$ for water):
\begin{equation}
\calF[n]=\Theta[n]-\Theta_{\mathrm{B}}.
\end{equation}

This leads to 
\begin{align}
\calF[\nofr] & =\kbT\int\left[\nofr\ln\left(\frac{\nofr}{\nb}\right)-\nofr+\nb\right]\dr\nonumber \\
 & +\int\nofr\Phi_{\mathrm{LJ}}(\boldsymbol{r})\dr+\Fexc[\nofr].\label{eq:F=00003DFid+fext+fexc}
\end{align}
The terms of the right-hand side of Eq.\ref{eq:F=00003DFid+fext+fexc}
corresponds to the usual decomposition\cite{evans_nature_1979,evans_density_2009,henderson_fundamentals_1992}
into an ideal term accounting for information entropy, an external
term accounting for the perturbation by the solute through its external
potential, and an excess term accounting for solvent-solvent correlations.
This last, excess term, can be rewritten without additional approximation
as 
\begin{align}
\Fexc[\nofr] & =-\frac{\kbT}{2}\iint\Delta\nofr c(r)\Delta n(\boldsymbol{r}^{\prime})\dr\dr^{\prime}+\Fcor\label{eq:Fexc}\\
 & =\FexcHNC+\Fcor,\nonumber 
\end{align}
where $r\equiv\left\Vert \boldsymbol{r}-\boldsymbol{r}^{\prime}\right\Vert $,
$\Delta\nofr\equiv\nofr-\nb$, and $c(r)$ is the direct correlation
function of the homogeneous reference fluid at density $\nb$. The
first term thus corresponds to a series expansion in density of $\Fexc$,
around the density of the HRF, truncated at second order. Truncated
information is put into an unknown bridge term, $\Fcor$. When $\Fcor=0$
, i.e. when we stick to the pure HRF approximation, Eq.\ref{eq:Fexc}
can be shown to correspond to the HNC approximation of integral equations\cite{hansen_theory_2006}.
It will thus be called the HNC functional below. We suppose now that
the correction can be expressed as a polynomial containing all terms
of orders higher than 2 in $\Delta n$. Eq.\ref{eq:Fexc} can be used
only if one knows the direct correlation function, $c(r)$. In this
article, we use an accurate direct correlation function of SPC/E water
computed by Belloni et al. according to the methods discussed in refs.~
\cite{puibasset_bridge_2012,belloni_efficient_2014}. The Fourier
transform of the direct correlation function, $\hat{c}$, is calculated
by 
\begin{equation}
\hat{c}\left(k\right)=\frac{\hat{h}_{000}\left(k\right)}{1+n_{B}\hat{h}_{000}\left(k\right)},
\end{equation}
where $\hat{h}_{000}\left(k\right)$ is the first rotational invariant
of the Fourier transform of the total correlation function $h$ calculated
by Puibassset and Belloni~\cite{puibasset_bridge_2012}. This function,
as well as all higher rotational invariant components are obtained
at short range by Monte Carlo sampling, and at higher range by integral
equation closures, so that small $k$ values are very accurate.

The HNC functional has proved to be good enough for studying solvation
in acetonitrile and in the Stockmayer fluid, but it exhibits wrong
behaviors when coming to water\cite{zhao_molecular_2011}. To improve
the description of water we have proposed, as several other authors
\cite{_fast_Wu_2014,oettel_integral_2005}, an hard sphere bridge
functional that consists in replacing all the unknown orders in $\Delta n$
of the molecular fluid by the known ones of a hard sphere fluid of
a diameter chosen on physical considerations\cite{levesque_scalar_2012}.
Such a correction does improve the solvation of small molecular solutes~\cite{zhao_new_2011,zhao_correction_2011,levesque_scalar_2012}.
However, the HNC functional or the functional with the hard sphere
bridge, HNC+HSB, are not able to reproduce to date the solvation of
hydrophobic solutes at both small and large length scales. This is
an important discrepancy that originates from the fact that water
at room conditions is close to liquid-gas coexistence and has a very
low pressure, a fact that is impossible to account for consistently
in HNC~\cite{oettel_integral_2005}.

We proposed recently a correction that imposes the essential physics\cite{jeanmairet_molecular_2013}.
It is based on the separation of the functional of Eq.\ref{eq:F=00003DFid+fext+fexc}
with the hard sphere correction in a short range and a long range
part. The long range part was then made compatible with the Van-der-Waals
theory of phase coexistence at long range in a spirit similar to the
Lum-Chandler-Weeks theory\cite{lum_hydrophobicity_1999}. It introduces
a coarse-grained density, similar in nature to weighted densities
at the core of fundamental measure theories for hard sphere fluids\cite{rosenfeld_free-energy_1989}.
We were then able to reproduce qualitatively the solvation of hydrophobic
solutes at all length-scales. The surface tension was found too high,
however, and the solvation structure of qualitative agreement only.
It should be noted that the key role of the pressure of the fluid
was not identified in this work: The pressure was consequently not
explicitly considered as a control parameter even if this correction
had an effect on the pressure. A functional that imposes the coexistence
and the right pressure of the fluid is thus presented here.

What we propose is an expression of $\Fcor$ that is cubic in $\Delta n$.
There are two main motivations to such an expression. (\emph{i}) First,
Evans et al\cite{evans_failure_1983} and later Rickayzen and collaborators\cite{rickayzen_integral_1984,powles_density_1988} showed that a series expansion of the
functional at the quadratic order is thermodynamically inconsistent
. In particular, the pressure
of the homogeneous reference fluid predicted by the theory can  be overestimated
by orders of magnitude. For instance, for water, the HNC functional
predicts a pressure of approximately 11450~bar instead of 1~bar.
(\emph{ii}) Also, Rickaysen proposed to add the simplest cubic term
to the series expansion of $\mathcal{F}_{\mathrm{exc}}$ in density,
and showed it to be sufficient to overcome the thermodynamic inconsistency.
Following Rickayzen's prescriptions, a ``three body'' bridge term
that is cubic in $\Delta n$, $\mathcal{F}_{3\mathrm{B}}$, is proposed.
We give arguments on the form that should have $\mathcal{F}_{3\mathrm{B}}$
for water, and how the addition of a physical constraint makes it
a single parameter functional.

Instead of the simple three body expression of Rickayzen based on
the overlap of hard bodies, we use here a rather different expression
that is motivated by the fact that in water, tetrahedral order due
to hydrogen bonding is lost in the HNC approximation and should be
reinforced. Note that if the structuration discussed below is particular
to water, the idea of including a three body term to improve the local
structuration given by the HNC functional is relevant for any solvent.

The three body functional should \textit{(i)} enforce thermodynamic
consistency, \textit{(ii)} give back the local order due to $N$-body
interactions missed so far ($N>2$), and \textit{(iii)} stay numerically
efficient since it is our long-term goal to compete with other implicit
methods like PCM (Polarizable Continuum Model)\cite{tomasi_quantum_2005} that are much cruder but
extremely useful. This last point may seem minor from a physical point
of view; Nevertheless, to compute $\mathcal{F}_{3\mathrm{B}}$, one
should integrate over the whole $\mathbb{R}^{9}$ instead of $\mathbb{R}^{6}$
(with convolutions) for HNC: if it is not built efficiently, then
it is useless. Consequently, in addition to physical motivation, the
analytical form of the three body functional introduced here must
allow efficient computation.

We start from the idea of the coarse-grained model of tetracoordinated
silicon by Stillinger and Weber\cite{stillinger_computer_1985,stillinger_erratum:_1986},
re-parameterized later for water by Molinero and Moore\cite{molinero_water_2009}.
Their idea relies on an harmonic penalty to non-tetrahedral oxygen-oxygen-oxygen
angles. In the MDFT framework, it leads to
\begin{align}
\beta\Fthree[n(\bm{r})] & =\frac{\lambda}{2}\int\Delta n(\bm{r}_{1})\biggl[\iint\Delta n(\bm{r}_{2})\Delta n(\bm{r}_{3})f(r_{12})f(r_{13})\nonumber \\
 & \times\left(\frac{\bm{r}_{12}\cdot\bm{r}_{13}}{r_{12}r_{13}}-\cos\theta_{0}\right)^{2}\mathrm{d}\bm{r}_{2}\mathrm{d}\bm{r}_{3}\biggr]\mathrm{d}\bm{r}_{1},\label{eq:F3B expression en r}
\end{align}
with $\beta$=($\kbT$)$^{-1}$. The dot product defines the cosine
of the angle between three space points, and the quadratic term enforces
a tetrahedral angle with $\theta_{0}=109.5\text{\textdegree}$. The
function $f$ tunes the range of the three body interaction. As a
source of \emph{local} structuration of the fluid, it must be short-ranged
and must vanish after few solvent radii, at distance $r_{\mathrm{max}}$.
We propose as Molinero and Moore 
\begin{equation}
f(r)=\begin{cases}
\exp\left(\frac{2}{3}\frac{r_{\mathrm{max}}}{\left(r-r_{\mathrm{max}}\right)}\right) & \text{if \ensuremath{r<r_{\mathrm{max}}}}\\
0 & \text{if \ensuremath{r\geq r_{\mathrm{max}}}}
\end{cases}.
\end{equation}
$\lambda$ is a dimensionless parameter modulating the strength of
this oriented-bond term (hydrogen bond in case of water). The excess
term in Eq.\ref{eq:F3B expression en r} is specific to a given fluid
and should thus be parameterized once for all for the sake of consistency.
We chose it so that one recovers the thermodynamic consistency and
the correct pressure of the bulk liquid.

The grand potential of a system of homogeneous fluid of volume $V$
and pressure $\mathrm{P}$ is equal, by definition, to $-\mathrm{P}V$.
It is $0$ in an empty system. Thus, one can deduce the pressure in
the reference fluid by evaluating the functional of Eq.\ref{eq:F=00003DFid+fext+fexc}
at zero density\cite{sergiievskyi_fast_2014}:

\begin{equation}
{\cal F}[n=0]=\Theta[n=0]-\Theta_{\mathrm{B}}=\mathrm{P}V,\label{eq:pressur=00003DOmega(O)}
\end{equation}
Using Eq.\ref{eq:pressur=00003DOmega(O)} for the functional without
the three body term we get,
\begin{equation}
\beta\mathrm{P_{HNC}}=\nb-\frac{\nb^{2}}{2}\bar{c}\label{eq:PHNC}
\end{equation}
with $\bar{c}=4\pi\int_{0}^{\infty}r^{2}c(r)\mathrm{d}r$. With the
three-body term:
\begin{equation}
\beta\mathrm{P_{3B}}=\nb-\frac{\nb^{2}}{2}\bar{c}+\frac{32n_{\mathrm{B}}^{3}}{9}\pi^{2}\lambda\left[\int_{0}^{\infty}f(r)r^{2}\mathrm{d}r\right]^{2}.\label{eq:P3B}
\end{equation}
With Eq.\ref{eq:PHNC} we find a pressure above 11450 bar for the
HNC functional. Eq.\ref{eq:P3B} is used to fix the parameter $\lambda$
to have the desired pressure for the bulk fluid, i.e., 1 bar for water
at room conditions. With this constraint, the three body functional
has only one parameter left: the range of the interaction, $r_{\mathrm{max}}$.
Molinero and Moore determined a parameter $\rmax=4.3$~\AA\  for
their model. We kept the freedom of slightly varying $\rmax$ around
this value. An optimum value is found for $4.2$~\AA. See below.

The direct computation of the three-body function of Eq.\ref{eq:F3B expression en r}
cannot be performed because it requires a triple nested integration
over the spacial coordinates. To accelerate the computation of this
term we rewrite Eq.\ref{eq:F3B expression en r} as: 
\begin{align}
\beta\Fthree[n(\bm{r})] & =\frac{\lambda}{2}\int\Delta n(\bm{r}_{1})\left(\sum_{\alpha,\beta\in\left\{ x,y,z\right\} }\bar{n}_{\alpha\beta}(\bm{r}_{1})^{2}+\cos^{2}\left(\theta_{0}\right)\bar{n}{}_{0}(\bm{r}_{1})^{2}-2\cos\left(\theta_{0}\right)\bar{\bm{n}}_{1}(\bm{r}_{1})\cdot\bar{\bm{n}}_{1}(\bm{r}_{1})\right)\mathrm{d}\bm{r}_{1}\label{eq:F3B_with_convol}
\end{align}
where 
\begin{equation}
\bar{n}_{\alpha\beta}(\bm{r}_{1})=\int f(r_{12})\frac{\alpha_{12}\beta_{12}}{r_{12}^{2}}\Delta n(\bm{r}_{2})\mathrm{d}\bm{r}_{2}\text{, \ \ensuremath{\alpha,\beta\in\left\{ x,y,z\right\} }}\label{eq:Falphabeta}
\end{equation}
\begin{equation}
\bar{\bm{n}}_{1}(\bm{r}_{1})=\int f(r_{12})\frac{\bm{r}_{12}}{r_{12}}\Delta n(\bm{r}_{2})\mathrm{d}\bm{r}_{2},\label{eq:bold}
\end{equation}
\begin{equation}
\bar{n}{}_{0}(\bm{r}_{1})=\int f(r_{12})\Delta n(\bm{r}_{2})\mathrm{d}\bm{r}_{2}.\label{eq:F0}
\end{equation}
It can be seen that~$\Fthree$ belongs to the general class of weighted
functionals with one scalar weighted density, one vectorial one, and
one second order, tensorial one.

The derivation of the equivalence between Eq.\ref{eq:F3B expression en r}
and Eq.\ref{eq:F3B_with_convol} as well as the first- and second-order
functional derivatives that may be needed for minimizing Eq.\ref{eq:F3B expression en r}
are given in supplementary information\cite{SeeSup}. Convolution products of Eqs.\ref{eq:Falphabeta},
\ref{eq:bold} and \ref{eq:F0} are evaluated efficiently in three
dimensions using fast Fourier transforms (FFT). We typically use cubic
boxes of $35^{3}$ \AA$^{3}$ with space discretized by 5 grid nodes
per $\textrm{\AA}$. Functional minimization of the total functional
is typically reached within 15 to 20 iterations in a few tens of minutes
on a single processor core at 2.4 GHz.

\section{Results and Discussion\label{sec:Results-and-Discussion}}

Our goal is to predict the hydration structure and free energy of
hydrophobic solutes from microscopic to macroscopic length scales.
Hydration free energies of nanometric solutes are proportional to
the surface of the solute. Since this behavior is due to the almost
zero pressure of liquid water at room conditions, it is of prime importance
to build a density functional that imposes the right pressure. First,
we describe the parameterization of Eq.\ref{eq:F3B expression en r}
for capturing both the liquid-gas coexistence and the correct pressure.
After that, we test the functional against hydration of various apolar
solutes.

\begin{figure}
\centering{}\includegraphics[width=85mm]{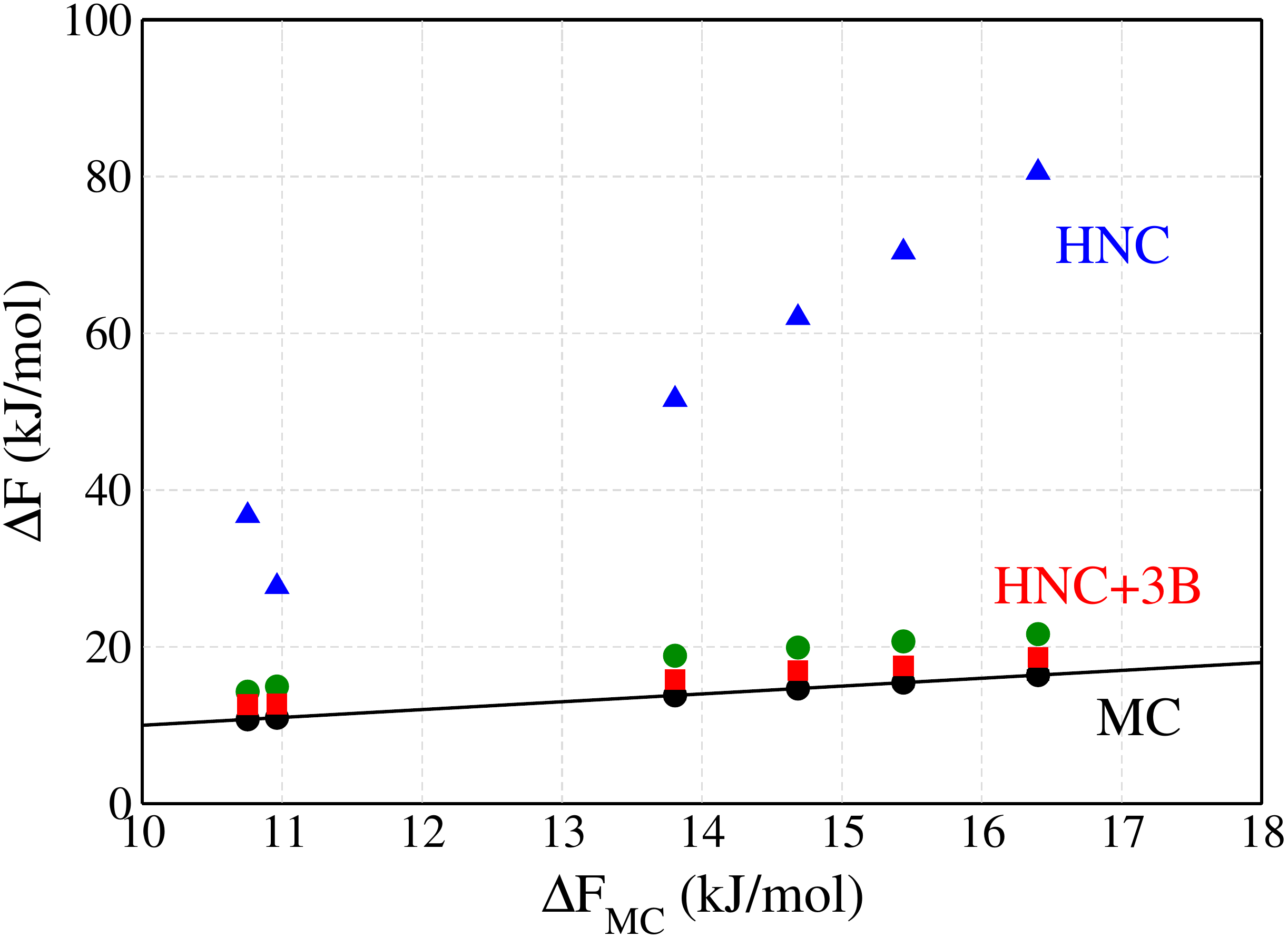}\protect\caption{Hydration free energies for the first six linear alkanes as calculated
with MDFT-HNC and MDFT-HNC+3B, compared to Monte Carlo simulations
by Ashbaugh et. al.~\cite{ashbaugh_hydration_1998}. $\rmax=4.2$~\AA\ 
(red squares) and $4.3$~\AA\  (green circles) are shown for MDFT-HNC+3B.\label{fig:Solvation-free-energy_alkanes}}
\end{figure}

To parametrize and test the three body term of Eq.\ref{eq:F3B expression en r}
we first study small molecular apolar solutes. In Fig.\ref{fig:Solvation-free-energy_alkanes},
we show the solvation free energies of small alkane chains as computed
by Monte Carlo simulations (MC)\cite{ashbaugh_hydration_1998} and
by MDFT-HNC or MDFT-HNC+3B with $\rmax=4.3$~\AA\ 
and $4.2$~\AA. Within MDFT-HNC, the error in solvation
free energy increases linearly with the size of the alkane, that is
its number of carbons, shown here from methane to hexane.
With the three-body
excess functional, the relative error of MDFT with respect to Monte
Carlo simulations is reduced by several orders.

We find that $\rmax=4.2$~\AA\ produces the optimal results, close to $4.3$~\AA\  for Molinero
and Moore and we stick to this value in the rest of the article. 
The remaining parameter $\lambda$
is chosen to impose the correct pressure in bulk water, $\mathrm{P}=1$~bar,
from Eq.\ref{eq:P3B}. One finds $\lambda=38$. We highlight that
since the pressure of the fluid is now correct, the pressure correction
term proposed by Sergiievskyi et al\cite{sergiievskyi_fast_2014}
is no longer required. 

\begin{figure}
\begin{centering}
\includegraphics[width=85mm]{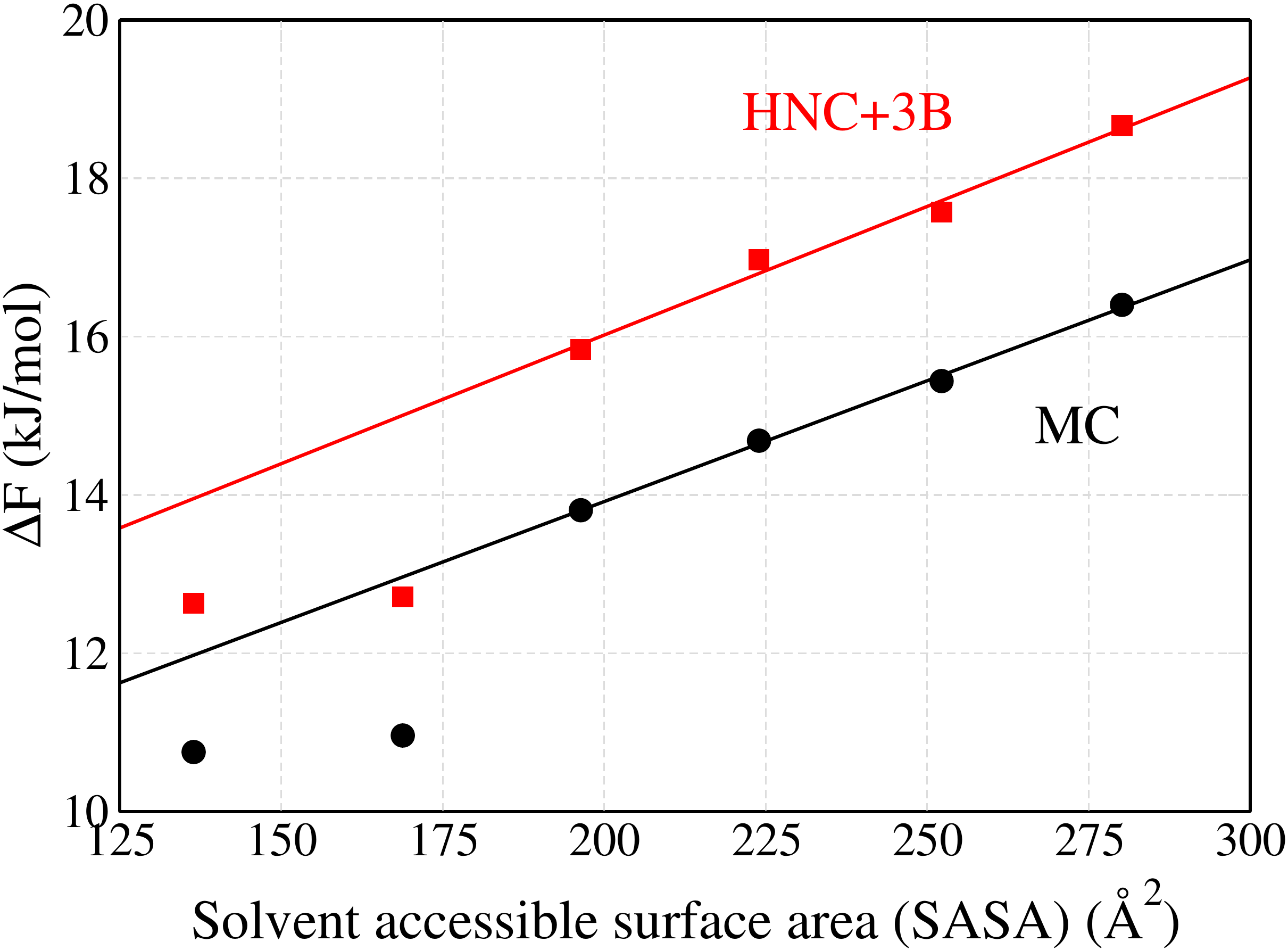}
\par\end{centering}

\protect\caption{Hydration free energy in SPC/E water for the first six linear alkanes
as a function of the solvent accessible surface area (SASA). Reference
results from Monte Carlo are plotted as black circles. MDFT results
with three-body corrections are in red squares. Linear regressions
based on propane, butane, pentane and hexane are also plotted.\label{fig:sasa}}
\end{figure}

The solvation free energy of $n$-alkanes into water is known to scale
linearly with the molecular surface area~\cite{bennaim_solvation_1984,ashbaugh_hydration_1998}:
\begin{equation}
\Delta \mathrm{F}=\gamma_{\textrm{m}}\mathcal{A}+b,\label{eq:sasa}
\end{equation}
with $\mathcal{A}$ the solute area, $\gamma_{\textrm{m}}$ the free
energy per microscopic surface area and $b$ an offset. Note that
$\gamma_{\textrm{m}}$ is a microscopic equivalent to a surface tension,
but is definitely different from the macroscopic liquid-gas surface
tension. Several definitions of $\mathcal{A}$ can be found in the litterature, that do not
change any conclusion therein: We will use the solvent accessible
surface area (SASA) of water in what follows, in order to be as comparable
as possible with the results by Ashbaugh et al.~\cite{ashbaugh_hydration_1998}.
The evolution
of the hydration free energy with respect to the solvent accessible
surface area is plotted in Fig.\ref{fig:sasa}. With the three-body excess functionnal,
we now find the anticipated linear dependancy. The values $\gamma_{\textrm{m}}$
and $b$ given by the linear regressions corresponding to equation~\ref{eq:sasa}
are given in Table~\ref{tab:Microscopic-equivalent-to}. The value
of $\gamma_{m}$ is in good agreement with both MC and experiments.
We get an offset of approximately one $\kbT$ with respect to MC.

\begin{table}
\begin{centering}
\begin{tabular}{|c|c|c|c|c|}
\hline 
 & HNC & HNC+3B & MD & Exp.\tabularnewline
\hline 
\hline 
$\gamma_{m}$ (J/(mol$\cdot$\AA$^{2}$)) & $340.59$ & $32.49$ & $30.53$ & $28.5$\tabularnewline
\hline 
$b$ (kJ/mol) & $-15.02$ & $9.52$ & $7.81$ & $2.51$\tabularnewline
\hline 
\end{tabular}
\par\end{centering}

\protect\caption{Microscopic equivalent to the surface tension and offset from MD~\cite{ashbaugh_hydration_1998},
from experiments\cite{bennaim_solvation_1984} and by MDFT-HNC and
MDFT-HNC+3B.\label{tab:Microscopic-equivalent-to}}
\end{table}

\begin{figure}
\centering{}\includegraphics[width=85mm]{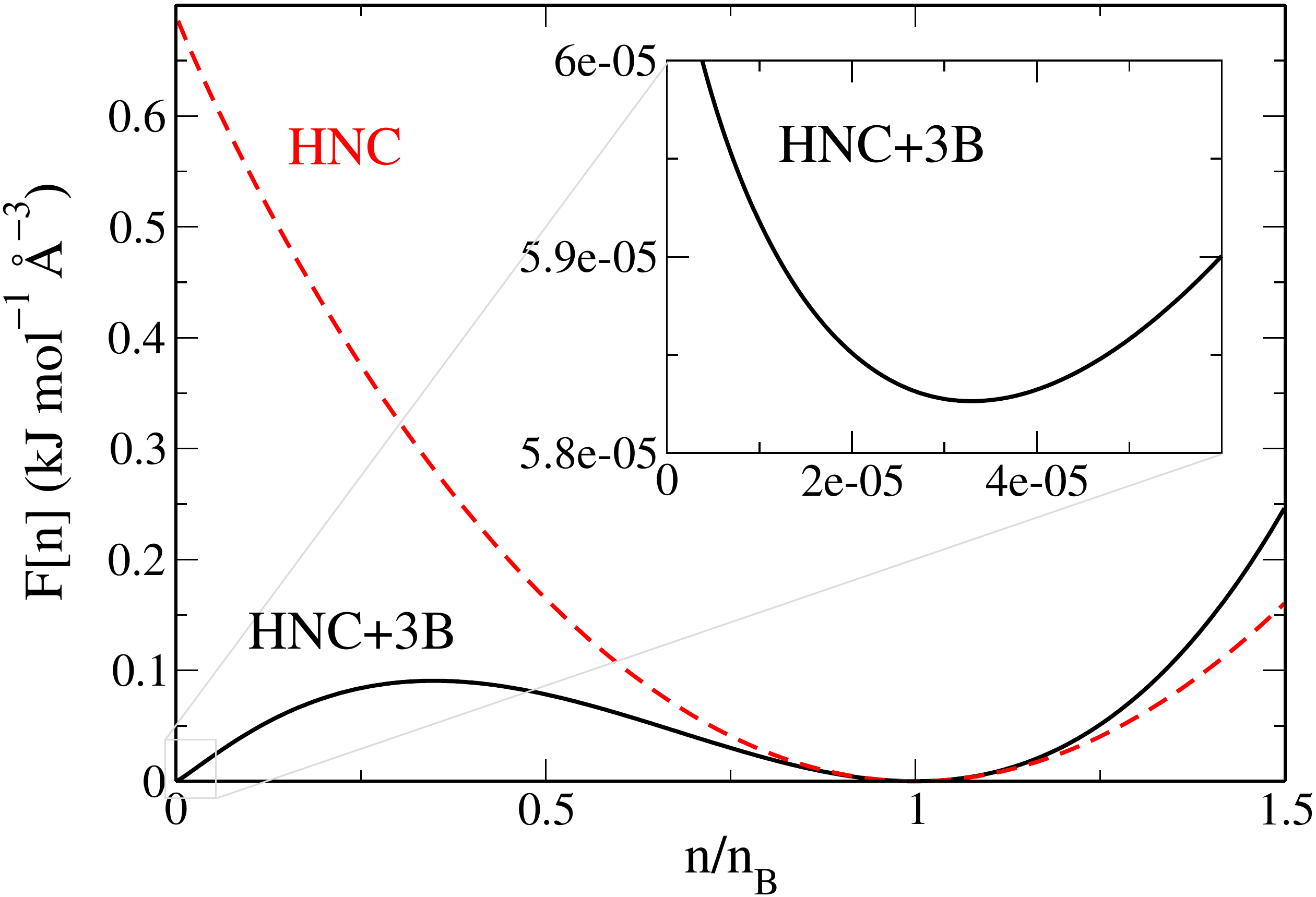}\protect\caption{Helmholtz free-energy of a homogeneous system of density $n$, see
Eq.~\ref{eq:F=00003DFid+fext+fexc}. $\nb$ is the reference density
one uses for the HNC functional. The insight is a focus on the first
local minimum of the three-body corrected functional, HNC+3B.\label{fig:wF3B}}
\end{figure}

Now that parameters are fixed once for all, we show in Fig.\ref{fig:wF3B}
the Helmholtz free energy of the homogenous systems as a function
of the density at 300~K, as computed with the MDFT-HNC functional
in dashed red and with the MDFT-HNC+3B functional in black. As discussed
above, no second phase can appear in the system described with the
MDFT-HNC functional since the free energy has only one minimum. Consequently, it
can not capture liquid-gas coexistence\cite{evans_failure_1983}. On the other hand, there are
two minima of the Helmholtz free energy for the functional that includes
the three-body bridge functional. A local minimum is found close to zero-density (``a gas phase'') with a free energy larger than the
one of the global minimum corresponding to the density of the reference
homogeneous fluid. The difference in Helmholtz free energy is of the
order of $6.0.10^{-5}$~kJ/\AA$^{3}$, the homogeneous water we
are describing is thus liquid and very close to liquid-gas coexistence.
This physical feature is a key \cite{dzubiella_competition_2004}
to predict the solvation structure of large hydrophobic solutes of
nanometer scale. 

To summarize: \textit{(i)} the cost in free energy per unit volume
for creating a cavity within the HNC (or HRF) formalism is several orders of
magnitude too high, in relation to its overestimation of the pressure,
\textit{(ii)} the bridge functional that we propose corrects both
the local order and the pressure, and it induces that the system is
close to coexistence. The cost for creating a cavity within the MDFT-HNC+3B
formalism is thus reduced to almost zero.

\begin{figure}
\begin{centering}
\includegraphics[width=85mm]{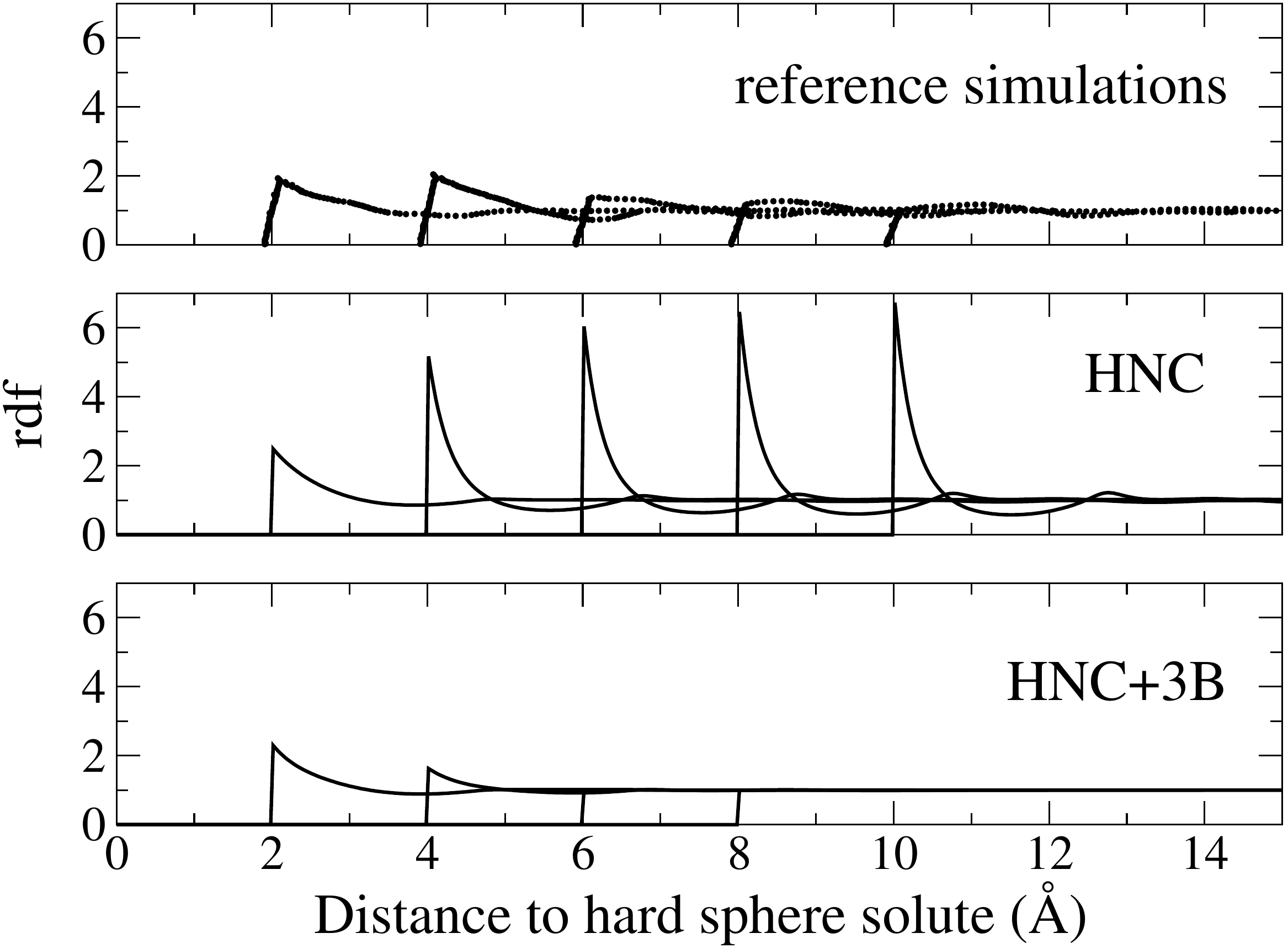}
\par\end{centering}

\protect\caption{Radial distribution function of a hard sphere solute of growing radius
in SPCE water at 300~K from reference Monte Carlo simulations~\cite{huang_hydrophobic_2002},
MDFT-HNC and MDFT-HNC+3B.\label{fig:Radial-distribution-function}}

\end{figure}

\begin{figure}
\centering{}\includegraphics[width=85mm]{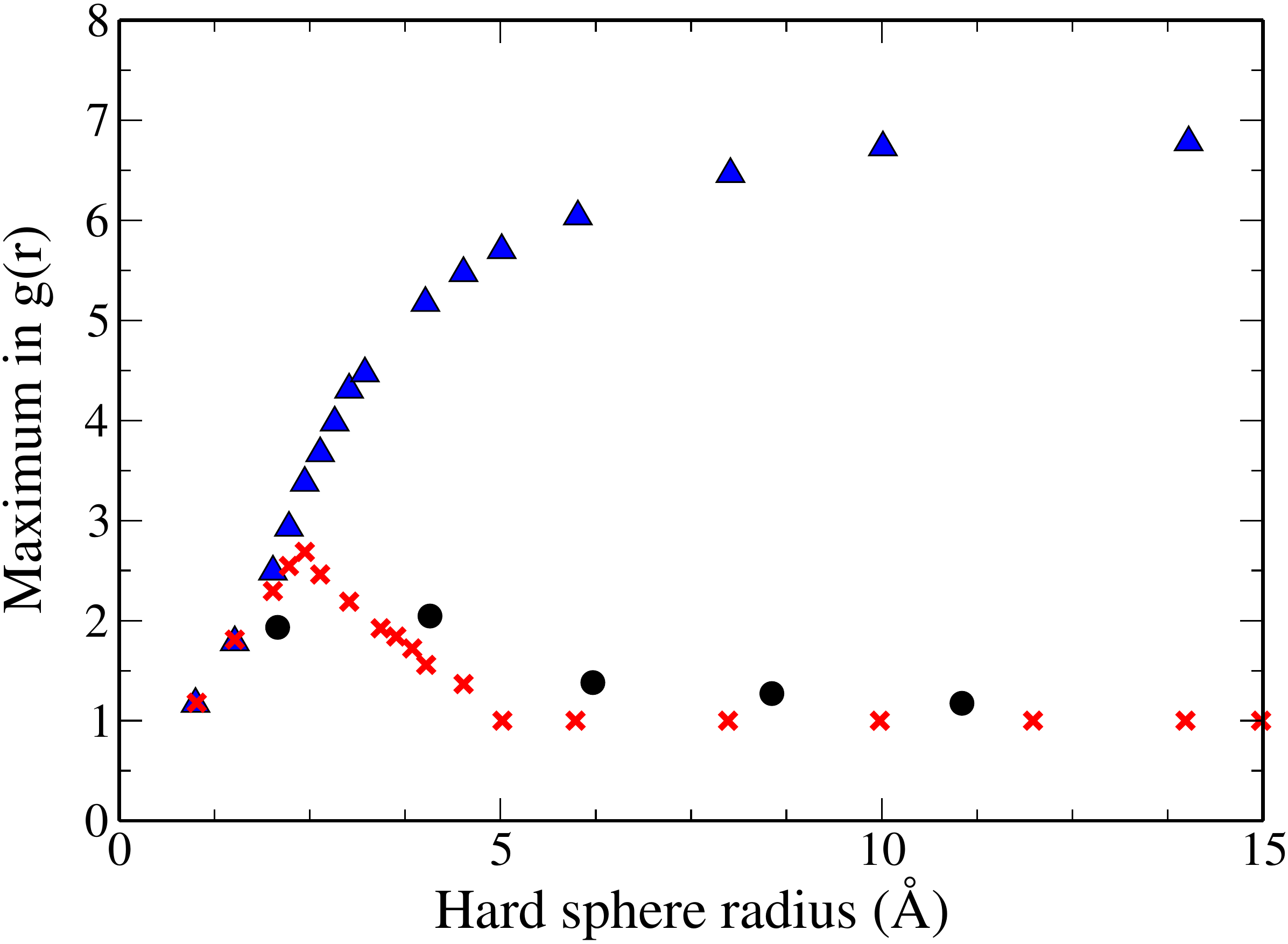}\protect\caption{Maxima of the radial distribution functions of hard sphere of different
radii $R$. The MC simulations results of Huang \textit{et al.}~\cite{huang_hydrophobic_2002} are
the black circles, the ones obtained by MDFT-HNC+3B are the red crosses
and the results of MDFT-HNC are the blue triangles. \label{fig:gSD}}
\end{figure}

\begin{figure}
\centering{}\includegraphics[width=85mm]{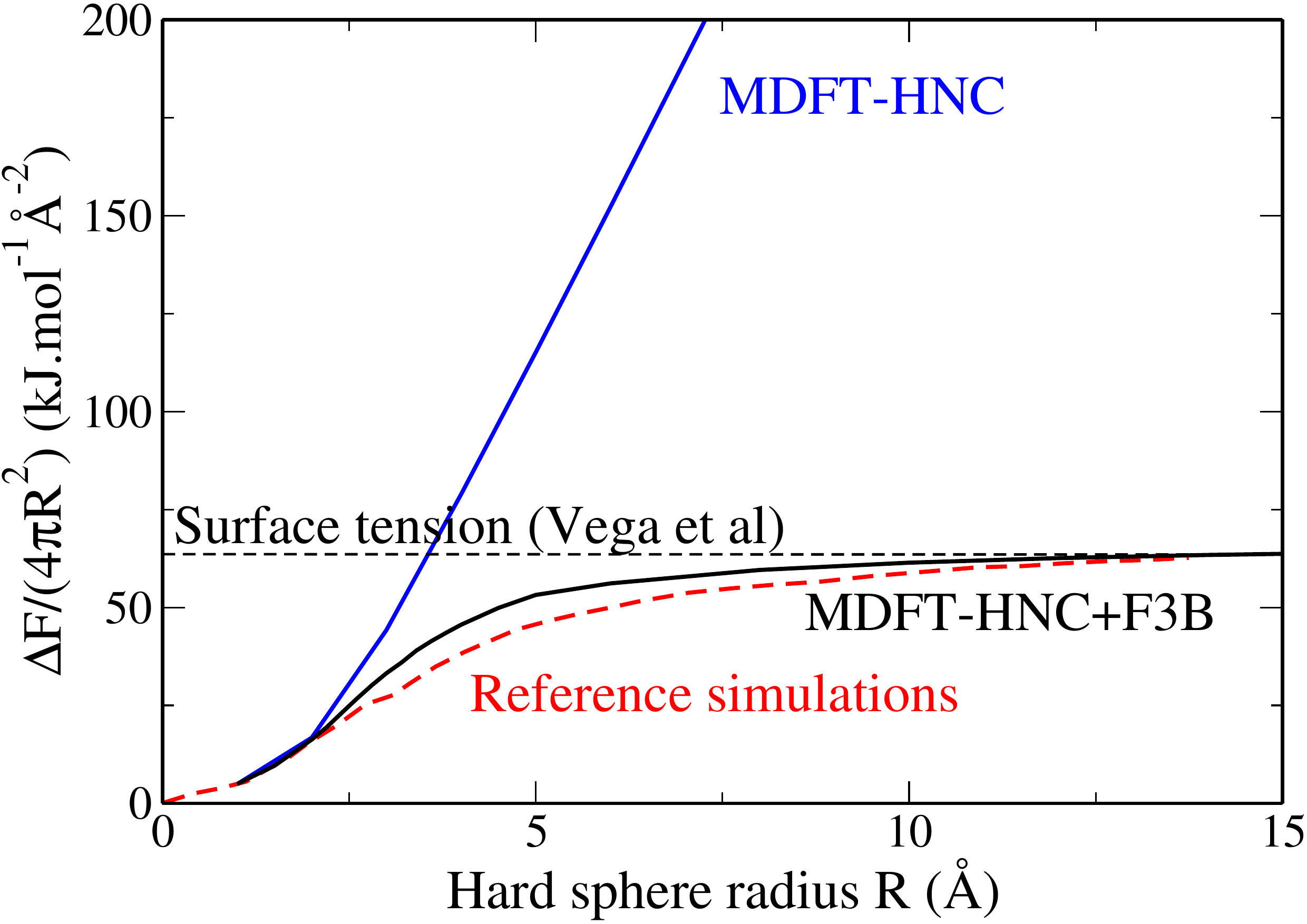}\protect\caption{Solvation free energy for hard spheres of different radii $\mathrm{R}$.
The value of the liquid-vapor surface tension of SPC/E water at 300
K estimated by Vega et al\cite{vega_surface_2007} (63.6 mJ/m$^{2}$)
is shown in dotted black line. MDFT-HNC+3B gets the right behavior:
two regimes at small then large HS radii, and the correct surface
tension.\label{fig:enerSD-1}}
\end{figure}

We now focus on the solvation of hydrophobic solutes of atomic to
nanoscale sizes. In their seminal works, Chandler and collaborators\cite{lum_hydrophobicity_1999,huang_hydrophobic_2002}
studied by Monte Carlo simulations the hydration of hard spheres whose
radii range from angstroms to nanometers. They observed a maximum
in height of the first peak of the hard sphere - water radial distribution
function at approximately 5~\AA. For radii larger than about 10~\AA,
they also observed a slow convergence toward a plateau for the surface
free energy. We compare the results by MDFT-HNC and MDFT-HNC+3B to
those of Huang \textit{et al}. in Fig.\ref{fig:Radial-distribution-function},
Fig.\ref{fig:gSD} and Fig.\ref{fig:enerSD-1}. One should keep in
mind that MDFT results are approximatively 1000~times faster than
explicit molecular dynamics or Monte Carlo simulations and that no
other implicit solvent methods besides the LCW theory is able to reproduce
these thermodynamic properties.

In Fig.\ref{fig:gSD}, we present the evolution of the height of the
first peak of the hard sphere (HS) - water radial distribution function
when the HS radius increases. This maximum corresponds to the most
probable distance of molecules of the first solvation shell to the
center of the hard sphere solute. The reference data by explicit methods
are given in black~\cite{huang_hydrophobic_2002}. This height exhibits
a peculiar maximum that is characteristic of the solvation of hydrophobic
solutes in water\cite{dzubiella_competition_2004}. It tends toward
unity for large radii. This behavior has been explained as follow:
for small radii, the solvent can reorganize around the solute without
losing solvent-solvent interactions, that is without losing too much
cohesion: The increase of the height of the peak is due to an increase
in packing of molecules at the surface of the sphere. For bigger radii,
the perturbation is too high to keep the local structure unchanged:
there is a loss of solvent-solvent interactions that has an energetic
cost that limits the accumulation of molecules at the surface of the
sphere and induces dewetting eventually. As a summary, when the perturbation
stays small compared to solvent cohesion, the packing increases around
the solute. Then, when the perturbation (the size of the solute) is
unfavorable compared to solvent cohesion, solvent molecules stand
back and the packing decreases. 

MDFT-HNC fails to reproduce this behavior: as depicted in Fig.\ref{fig:wF3B}
there is no possible change of regime for the fluid. With the three
body term, this change of regime can be found if the perturbation
is able to make the fluid reach a state close to the second minimum.
This is confirmed by Fig.\ref{fig:gSD}, where MDFT-HNC+3B is in qualitative
agreement with reference all atom simulations: The maximum of the
radial distribution function is obtained around $2.5\ \textrm{\AA}$,
which is reasonable a value. The decrease is, however, too fast.

In Fig.\ref{fig:enerSD-1} we plot the solvation free energy of
HS solutes per surface unit. We compile therein the results
by MDFT-HNC, MDFT-HNC+3B and once again the reference all atom Monte
Carlo simulations. MC shows a linear increase of the surface free
energy for small radii, followed by a transition state, then followed
by a plateau. This asymptotic value, reached at large HS radii corresponds
to the surface tension of the fluid. At this regime, the solvation
is thus driven by a sole surface term that corresponds at the microscopic
level to the case where the loss of interaction between solvent particles
is the prominent energetic term. MDFT-HNC is in agreement with the
simulations only for very small radius (below $2.5\ \textrm{\AA}$)
but does not reproduce the plateau for bigger radius ($>10\ \textrm{\AA}$),
this is consistent with the structural results, the transition between
the two regimes is missed. Again, MDFT-HNC+3B is in good agreement
with the simulations and the surface tension of SPC/E 
 water estimated by Vega et al \cite{vega_surface_2007} is recovered.

We can thus relate the decay of the maximum of the radial distribution
function in Fig.\ref{fig:gSD} and the convergence to the plateau
in Fig.\ref{fig:enerSD-1}. For structural and energetic properties,
Monte Carlo simulations show a smooth transition between the two regimes
described above, while MDFT-HNC+3B sharpens the transition: The three
body term exacerbates the importance of the loss of attraction between
solvent molecules.

To conclude this section, \textit{(i)} the structural properties obtained with
MDFT-HNC+3B are improved with respect to MDFT-HNC since we recover
the change in regime observed in MC at least qualitatively; \textit{(ii)}
this is also true for the the solvation free energy and this represents
a considerable progress since MDFT-HNC predicts the wrong quantitative
behavior. \textit{(iii)} The surface tension, that is related to the
height of the saddle point in the free energy curve of Fig.\ref{fig:wF3B},
is correctly reproduced by MDFT-HNC+3B even though this is not explicitly
controlled.

\section{Conclusions}

In this paper, we propose to go beyond the usual quadratic expansion
of the Gibbs free energy (or equivalently of the excess functional)
around the homogeneous reference fluid within the molecular density functional theory framework. We thus go beyond the HNC approximation. MDFT-HNC+3B imposes a second local
minimum to the Gibbs free energy of the system at low fluid density.
The bridge functional that was proposed \textit{(i)} enforces the
tetrahedral order of water, \textit{(ii)} recovers the close coexistence
between gas and liquid states and their surface tension, and \textit{(iii)}
is consistent with the experimental pressure of the fluid. It introduces
one empirical parameter that we fix to parameterize over the solvation
free energy of the first linear alkanes. It recovers the reference
results of explicit simulations with a systematic offset of order
$\kbT$. That is close to chemical accuracy, and is a clear improvement
over MDFT-HNC.

One advantage of this additional term with respect to previous work\cite{jeanmairet_molecular_2013}
is that \textit{(i)} it has a single empirical parameter, \textit{(ii)}
it does not require additional fields like coarse-grained densities,
and \textit{(iii) }it makes the theory thermodynamically consistent.

This bridge functional was used to study the solvation free energy
of hard spheres whose radii range from angstroms to nanometers. Unlike
MDFT-HNC, MDFT-HNC+3B recovers the change of regime between a solvation
governed by distortion of the solvent structure and a solvation governed
by a complete reorganization of the solvent. The free energy of solvation
and the surface tension are correct. Nevertheless, the transition
stage is too sharp.

Indeed, this points out the necessity of further improvements of the functional. Other thermodynamic  properties pertinent to hydrophobic solvation, such as entropy, enthalpy, partial molar volumes, temperature dependence, remain to be carefully tested too.

Numerical efficiency is the very essence of implicit methods like
MDFT. The bridge functional introduced therein would cause a dramatic
increase of the numerical cost without its rewriting in terms of fast
Fourier transforms. This is an important result of this article. The
numerical cost increase is at this stage of one order of magnitude
only with respect to MDFT-HNC. MDFT-HNC+3B is still two to three orders
of magnitudes faster than explicit simulations.

Finally the solutes studied here are all apolar and neutral, for the
sake of clarity and pedagogy. The three-body functional is built to
account for short-range tetrahedral order in the solvent. It is similar
in spirit to solute-solvent corrections that were introduced previously
in the group to describe ions and H-bonded polar solutes~\cite{jeanmairet_molecular_2014,jeanmairet_molecular_2013-1,jeanmairet_molecular_2013}.
We think this will lead to a consistent functional for water, valid
for both hydrophobic and hydrophilic interactions.

\begin{acknowledgments}
The authors thank Luc Belloni for providing a very accurate direct
correlation function of water and for fruitful discussions. Bob
Evans is greatly acknowledged for his input at the basis of this work
and for fruitful and delightful discussions.
\end{acknowledgments}
\bibliographystyle{unsrt}


\end{document}